\documentclass[prb,twocolumn,aps,superscriptaddress,showpacs]{revtex4-2}

\usepackage{amsmath,amssymb}
\usepackage{graphicx}
\usepackage{xcolor}
\usepackage{graphicx}
\DeclareGraphicsExtensions{.pdf,.eps,.png,.jpg} \graphicspath{{./figs/}}
\usepackage{dcolumn}
\usepackage{bm}
\usepackage{tabularx}
\usepackage{upgreek}
\usepackage{multirow}
\usepackage{epstopdf}
\usepackage{gensymb}
\usepackage{inputenc}
\usepackage{listings}

\usepackage{amsmath}%
\usepackage{amsfonts}%
\usepackage{amssymb}%
\usepackage[T1]{fontenc}
\usepackage{graphicx}
\usepackage{epstopdf}
\usepackage{color}
\usepackage[english]{babel}
\usepackage{natbib}
\usepackage[colorlinks=true, citecolor=blue, linkcolor=blue, urlcolor=blue]{hyperref}
\usepackage{placeins}
\usepackage{booktabs}
\usepackage{enumitem}
\usepackage{subfigure}
\usepackage{multirow}
\usepackage{braket}
\usepackage{svg}
\usepackage{dcolumn}
\usepackage{bm}
\usepackage{siunitx}
\usepackage{mathtools}
\usepackage{soul}

\newcommand{\beginsupplement}{%
  \setcounter{figure}{0}   \renewcommand{\thefigure}{S\arabic{figure}}
  \setcounter{table}{0}    \renewcommand{\thetable}{S\arabic{table}}
  \setcounter{equation}{0} \renewcommand{\theequation}{S\arabic{equation}}}

\begin{document}

\title{
Quantum disordered ground state and relative proximity to an exactly solvable model in the frustrated magnet CeMgAl$_{11}$O$_{19}$}

\author{G. Bastien}
\affiliation{Charles University, Faculty of Mathematics and Physics, Department of Condensed Matter Physics, Prague, Czech Republic}
\author{A. Eli\'a\v{s}}
\affiliation{Charles University, Faculty of Mathematics and Physics, Department of Condensed Matter Physics, Prague, Czech Republic}
\author{V. Anderle}
\affiliation{Charles University, Faculty of Mathematics and Physics, Department of Condensed Matter Physics, Prague, Czech Republic}
\author{A. Kancko}
\affiliation{Charles University, Faculty of Mathematics and Physics, Department of Condensed Matter Physics, Prague, Czech Republic}
\author{C. A. Corr\^ea}
\affiliation{Institute of Physics of the Czech Academy of Sciences, Department of Structure Analysis, Prague, Czech Republic}
\author{S. Kumar}
\affiliation{Charles University, Faculty of Mathematics and Physics, Department of Condensed Matter Physics, Prague, Czech Republic}
\affiliation{Faculty of Physics and Astronomy, Department of Experimental Physics of Condensed Phase, Adam Mickiewicz University, Pozna\'n, Poland}
\author{P. Proschek}
\affiliation{Charles University, Faculty of Mathematics and Physics, Department of Condensed Matter Physics, Prague, Czech Republic}
\author{J. Prokle\v{s}ka}
\affiliation{Charles University, Faculty of Mathematics and Physics, Department of Condensed Matter Physics, Prague, Czech Republic}
\author{L. N\'adhern\'y}
\affiliation{Department of Inorganic Chemistry, University of Chemistry and Technology Prague, Czech Republic}
\author{D. Sedmidubsk\'y}
\affiliation{Department of Inorganic Chemistry, University of Chemistry and Technology Prague, Czech Republic}
\author{T. Treu}
\affiliation{Experimental Physics VI, Center for Electronic Correlations and Magnetism, Institute of Physics,
University of Augsburg, 86159 Augsburg, Germany}
\author{P. Gegenwart}
\affiliation{Experimental Physics VI, Center for Electronic Correlations and Magnetism, Institute of Physics,
University of Augsburg, 86159 Augsburg, Germany}
\author{M. Kratochv\'ilov\'a}
\affiliation{Charles University, Faculty of Mathematics and Physics, Department of Condensed Matter Physics, Prague, Czech Republic}
\author{M. \v{Z}onda}
\affiliation{Charles University, Faculty of Mathematics and Physics, Department of Condensed Matter Physics, Prague, Czech Republic}
\author{R. H. Colman}
\affiliation{Charles University, Faculty of Mathematics and Physics, Department of Condensed Matter Physics, Prague, Czech Republic}

\date{\today}

\begin{abstract}
The magnetic properties of the triangular magnet CeMgAl$_{11}$O$_{19}$ were investigated by magnetization and specific heat measurements down to $T=0.03\,$K on single crystals grown by the floating zone method. The formation of effective spins $S_\mathrm{eff}= 1/2$ below $T < 10$\,K was confirmed both by DFT calculations and specific heat measurements. No magnetic order was found down to $T=0.03\,$K despite the formation of magnetic correlations observed in specific heat. The measured magnetization was compared with DMRG computation and  their agreement supports the proposal of a strongly anisotropic  magnetic interaction antiferromagnetically  coupling the spin components in the $ab$ plane and ferromagnetically coupling the spin component along the $c$ axis. However, our quantitative study of the magnetization indicates a weaker proximity to quantum criticality between ferromagnetism and antiferromagnetism than the previous inelastic neutron scattering study. Finally, we propose that the absence of magnetic order in CeMgAl$_{11}$O$_{19}$ would most probably be related to the structural disorder revealed by single-crystal X-ray diffraction.
\end{abstract}


\maketitle


The research on quantum magnetism is motivated by the possible formation of magnetically disordered states where quantum entanglement plays a role such as quantum spin liquid (QSL) states~\cite{Balents2010, Savary2016, Shaginyan2020, Li2020}, valence bond solids~\cite{ Norman2020, Pustogow2022} and random singlet (RS) states~\cite{Li2017, Kimchi2018}. 
Over the past decade, Kitaev magnets on honeycomb lattices have attracted considerable attention due to the analytic prediction of a quantum spin liquid state at the limit of uniaxial magnetic interactions~\cite{Takagi2019, Trebst2022}. Interestingly, the $S=1/2$ triangular lattice antiferromagnet (TLAF) also reaches a point where it becomes exactly solvable and a spin liquid state is predicted~\cite{Momoi1992, Gao2024}.   

We consider the $S=1/2$ TLAF with XXZ anisotropy and without any off-diagonal coupling~\cite{Miyashita1986,Momoi1992,Yamamoto2014, Sellmann2015,Gao2024} :

\begin{equation}
\hat{H}=\sum_{<i,j>}{J_\perp(S_i^xS_j^x+S_i^yS_j^y)+J_zS_i^zS_j^z} 
\label{H}
\end{equation}

\noindent In the case of an anisotropic interaction antiferromagnetically coupling spin components in the $ab$ plane ($J_\perp>0$), and ferromagnetically coupling the spin component along the $c$ axis ($J_z<0$), this Hamiltonian harbors a quantum critical point (QCP) at $J_z/J_\perp=-0.5$. The QCP separates a 120$^\circ$ coplanar antiferromagnetic order from a ferromagnetic order~\cite{Momoi1992,Gao2024}.  At the QCP, the problem is exactly solvable and it harbors a spin liquid ground state describable as a superposition of various umbrella states~\cite{Miyashita1986,Momoi1992, Pal2021, Gao2024}.

Ce-based TLAF are promising materials for the search of real materials in proximity to this QCP, since most of them harbor effective spins $S_\mathrm{eff}=1/2$ at low temperature~\cite{Liu2016, Bastien2020, Xie2024,Ashtar2025,Cao2025} and strongly anisotropic magnetic interactions induced by the strong spin-orbit coupling~\cite{Bastien2020, Kulbakov2021, Avdoshenko2022,Xie2024a, Gao2024}. Recently, the TLAF CeMgAl$_{11}$O$_{19}$ was proposed to be located close to the QCP of the XXZ Hamiltonian~\ref{H} from inelastic neutron scattering (INS) experiments~\cite{Gao2024}. CeMgAl$_{11}$O$_{19}$ crystallizes in the magnetoplumbite structure (space group $P6_3/mmc$)~\cite{Verstegen1974, Haberey1981}, which is characterized by a large interplane separation between the triangular magnetic layers~\cite{Ashtar2019, Bastien2024}. The presence of horizontal and vertical mirror planes in this crystal structure cancels off-diagonal symmetric exchange, whereas they play a major role in other Ce-based TLAFs such as KCeS$_2$~\cite{Avdoshenko2022} and CsCeSe$_2$~\cite{Xie2024a}. Importantly, an absence of antiferromagnetic order in CeMgAl$_{11}$O$_{19}$ down to $T=0.05\,$K was reported from specific heat measurements~\cite{Gao2024, Cao2025}, neutron diffraction~\cite{Gao2024} and muon spectroscopy~\cite{Cao2025}. In addition, INS experiments pointed out an excitation continuum~\cite{Gao2024}. Therefore, the ground state of CeMgAl$_{11}$O$_{19}$ was proposed to be a quantum spin liquid induced by the proximity to the QCP of the XXZ Hamiltonian~\cite{Gao2024}.

In this letter, we report a single crystal study of CeMgAl$_{11}$O$_{19}$ confirming the formation of $S_\mathrm{eff}=1/2$ effective spins at low temperature and the absence of magnetic order down to $T=0.03$\,K. Considering magnetization as the appropriate probe to quantify the proximity to ferromagnetic quantum criticality, we combined magnetization measurements at temperature down to $T=0.03\,$K with density matrix renormalization group (DMRG) calculation. Our data confirm the proposal of strongly anisotropic magnetic interactions but it indicates a weaker proximity to  ferromagnetic quantum criticality than previously proposed from INS~\cite{Gao2024}. These results exclude the scenario of a QSL state induced mainly by magnetic quantum criticality and they point out the important role of structural disorder.

\begin{figure}
\begin{center}
\includegraphics[width=0.9\linewidth]{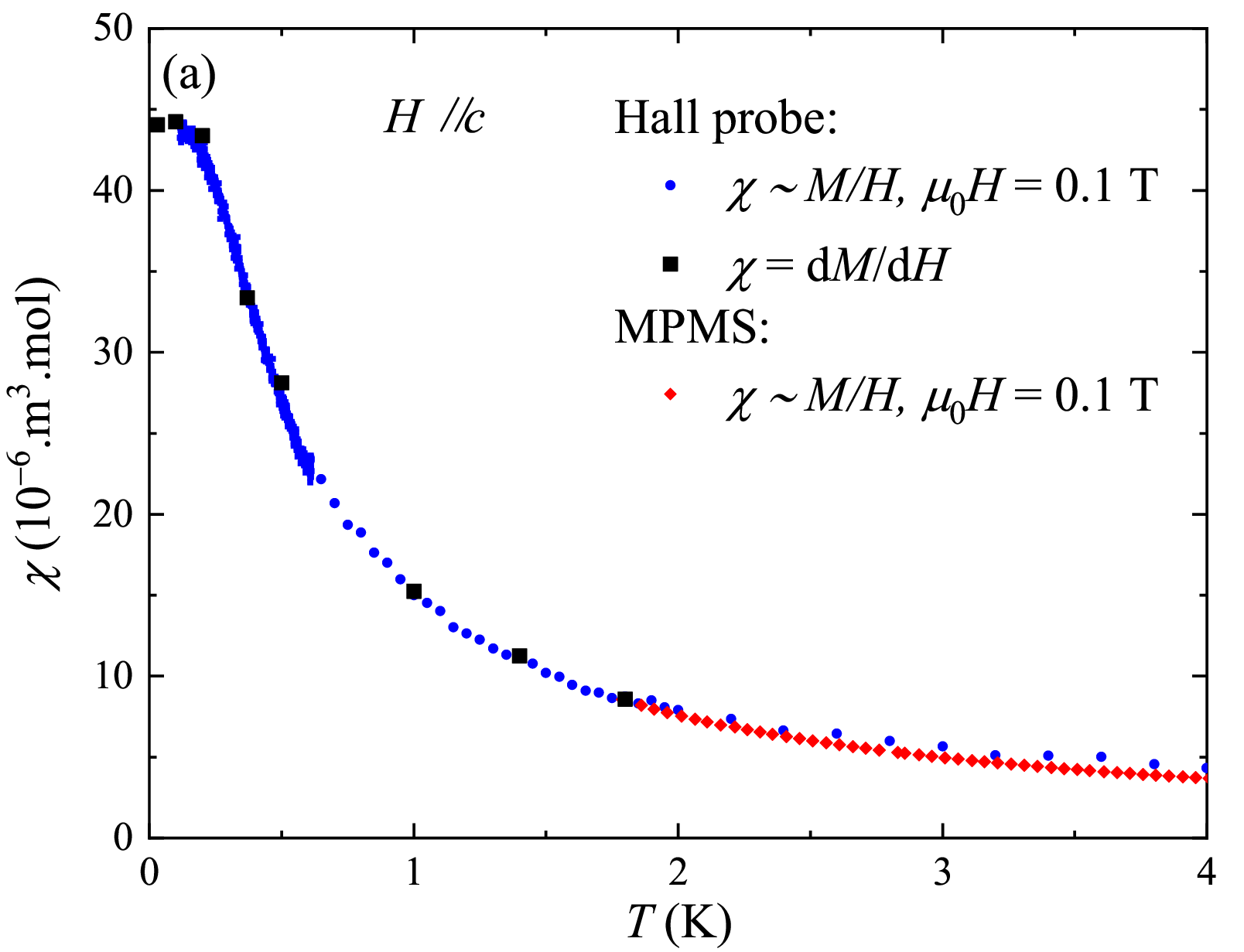}
\includegraphics[width=0.9\linewidth]{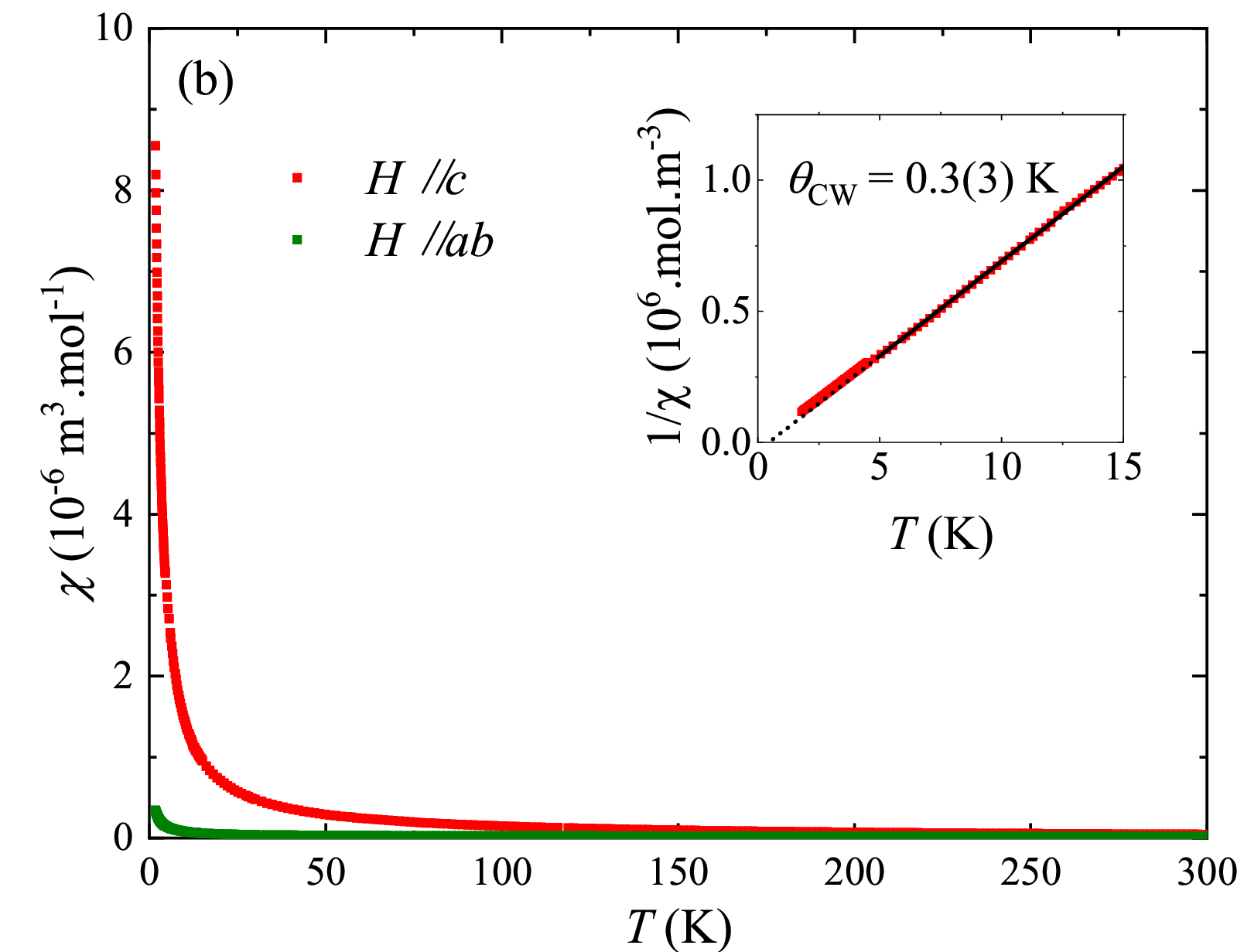}
\caption{(a) Magnetic susceptibility of CeMgAl$_{11}$O$_{19}$ along the easy magnetization axis $c$ as a function of temperature combining data from the MPMS magnetometer and the Hall sensor. Data point obtained upon temperature scan at a magnetic field of $\mu_\mathrm{0}H=0.1\,$T (discs) are confirmed by direct measurement of the derivative $\partial M/\partial H$ (black squares) (b) Magnetic susceptibility of CeMgAl$_{11}$O$_{19}$ under magnetic field applied both along the $c$ axis and in the $ab$ plane. The inset shows the Curie-Weiss fit in the temperature interval 5\,K$< T <$\,15\,K for magnetic field applied along the $c$ axis.}
    \label{MvsT} 
\end{center}
\end{figure}


The polycrystalline precursor for the crystal growth was first synthesized by sol-gel methods as described in Refs.~\cite{Dolezal2019, Nadherny2020}. Then single crystals were grown by the floating zone method in an optical furnace in air as described in Ref.~\cite{Kumar2025}. Powder X-ray diffraction revealed that our ingots contained grains of two different materials, the hexaaluminate CeMgAl$_{11}$O$_{19}$ and the spinel Mg$_{x}$Al$_{8/3-2x/3}$O$_{4}$. X-ray fluoresence spectroscopic mapping showed that these two phases are well separated and allowed confirmation of the composition of the CeMgAl$_{11}$O$_{19}$ phase (see Supplemental Materials~\cite{sup}).

Magnetization measurements at temperature down to $T = 1.8\,$K were performed using a MPMS magnetometer and a PPMS with VSM option, from Quantum Design. These measurements were extended down to $T = 0.03\,$K in a Triton dilution refrigerator  using the Hall probe method~\cite{Flanders1985, Bastien2024}. Magnetization measurements were carried out on a nearly cubic crystal with an estimated average demagnetization factor of $N\approx0.32$~\cite{Aharoni1998}. The specific heat was measured both by the relaxation method in the PPMS and the Triton dilution refrigerator and by adiabatic demagnetization refrigeration (ADR). The measurement by ADR involved the self-cooling of the sample from 1.8\,K to 81\,mK by demagnetization from 14 T to zero applied field, on a powdered sample. Constant heat flux on warming back to 1.8 K allowed calculation of the specific heat, as detailed in Refs.~\cite{Tokiwa2021, Treu2024}. 


Single-crystal diffraction confirms the magnetoplumbite crystal structure with the space group $P6_3/mmc$~\cite{Verstegen1974, Gao2024, Cao2025}. Experimental and structural details are given in the Supplemental Materials~\cite{sup, crysalis, Clark1995, Momma2008, Palatinus2007, Petricek2023}. The diffraction results unveil three sources of structural disorder. The Al(5) ion is distributed between two off-centered positions in the oxygen bipyramid as commonly observed in the magnetoplumbite structure~\cite{Abrahams1987, Bastien2024a, Gao2024, Cao2025,  Kumar2025,  Kumar2025a}. The present data also confirm that the Mg$^{2+}$ ions preferably sit at the Al (3) site as previously reported in the LnMgAl$_{11}$O$_{19}$ (Ln = La-Gd) series~\cite{Gasperin1984, Abrahams1987, Gao2024}. The best fit was achieved with 10 \% deficiency on the Ce site. This value is similar to the 7 \% deficiency previously reported in crystals grown in a reducing atmosphere~\cite{Gao2024}.

In order to test theoretically the description of CeMgAl$_{11}$O$_{19}$ as an effective spin $S_\mathrm{eff}=1/2$ magnet, we calculated the crystal electric field (CEF) energy
levels of the ground state multiplets of Ce$^{3+}$. These calculations were performed using the recently developed method based on a combination of DFT plane wave code using \emph{VASP}~\cite{Kresse1996} and Wien-2k. program~\cite{Blaha2001, Perdew1996,Kunes2010, Mostofi2008} (full computational details are given in Supplemental Materials~\cite{sup}). 
The crystal field parameters were defined from the Wannier transformed local Hamiltonian $\hat{H}_{loc}$
by expanding it into a series of spherical tensor operators $\hat{C}_q^k$

\begin{equation}\label{Hloc}
\begin{split}
 \hat{H}_{loc} = \sum_{k-2,4,6} \sum_{q=-k}^k B_q^k \hat{C}_q^k,\\
 B_0^2 = 52.16\,\textrm{meV},\ B_0^4 = 19.68\,\textrm{meV},\\ B_0^6 = -154.86,\,
 B_6^6 = B_{-6}^6 = -112.68\,\textrm{meV},
\end{split} 
\end{equation}

\noindent  
Solving the crystal field Hamiltonian
for $C_{3h}$ local symmetry of Ce$^{3+}$, we end up with
a diagonal matrix with the non-zero elements being equal to eigenvalues of three Kramers doublets

\begin{equation*}
\begin{split}
 E_{\pm5/2} = -\frac{2}{7} B_0^2 + \frac{1}{21} B_0^4 = -13.96\, \textrm{meV}, \\ 
 E_{\pm3/2} = \frac{2}{35} B_0^2 - \frac{1}{7} B_0^4 = +0.17\, \textrm{meV}, \\
 E_{\pm1/2} = \frac{8}{35} B_0^2 + \frac{2}{21} B_0^4 = +13.80\, \textrm{meV} .
\end{split} 
\end{equation*}

These results imply the quasi-exclusive selection of the $J_z=\pm 5/2$ magnetic state at temperatures $T\ll(E_{\pm5/2}-E_{\pm3/2})/k=164\,$K. Therefore, it validates previous interpretations of the magnetism of CeMgAl$_{11}$O$_{19}$ below $T=10$\,K as an effective spin $S_\mathrm{eff}=1/2$ magnet \cite{Gao2024,Cao2025}. The CEF scheme was further validated from the comparison of the  temperature dependence of the effective moment with experimental data as shown in the Supplemental Material~\cite{sup}.

The magnetic susceptibility of CeMgAl$_{11}$O$_{19}$ as a function of temperature down to $T=0.03\,$K is shown in Fig.~\ref{MvsT}(a). It is characterized by a strong anisotropy with the $c$ axis as the easy magnetization axis. Below $T=0.2\,$K, the DC magnetic susceptibility shows a plateau, in agreement with the AC magnetic susceptibility recently reported~\cite{Cao2025}. A Curie-Weiss fit performed in the temperature interval 5\,K$< T<$15\,K gives a Curie-Weiss  temperature of $\theta_\mathrm{CW}=0.3\pm0.3\,$K. The positive value hints at ferromagnetic interactions in agreement with the proposal of a magnetic interaction ferromagnetically coupling the $S^z$ spin component~\cite{Gao2024}.

\begin{figure}
\begin{center}
\includegraphics[width=0.9\linewidth]{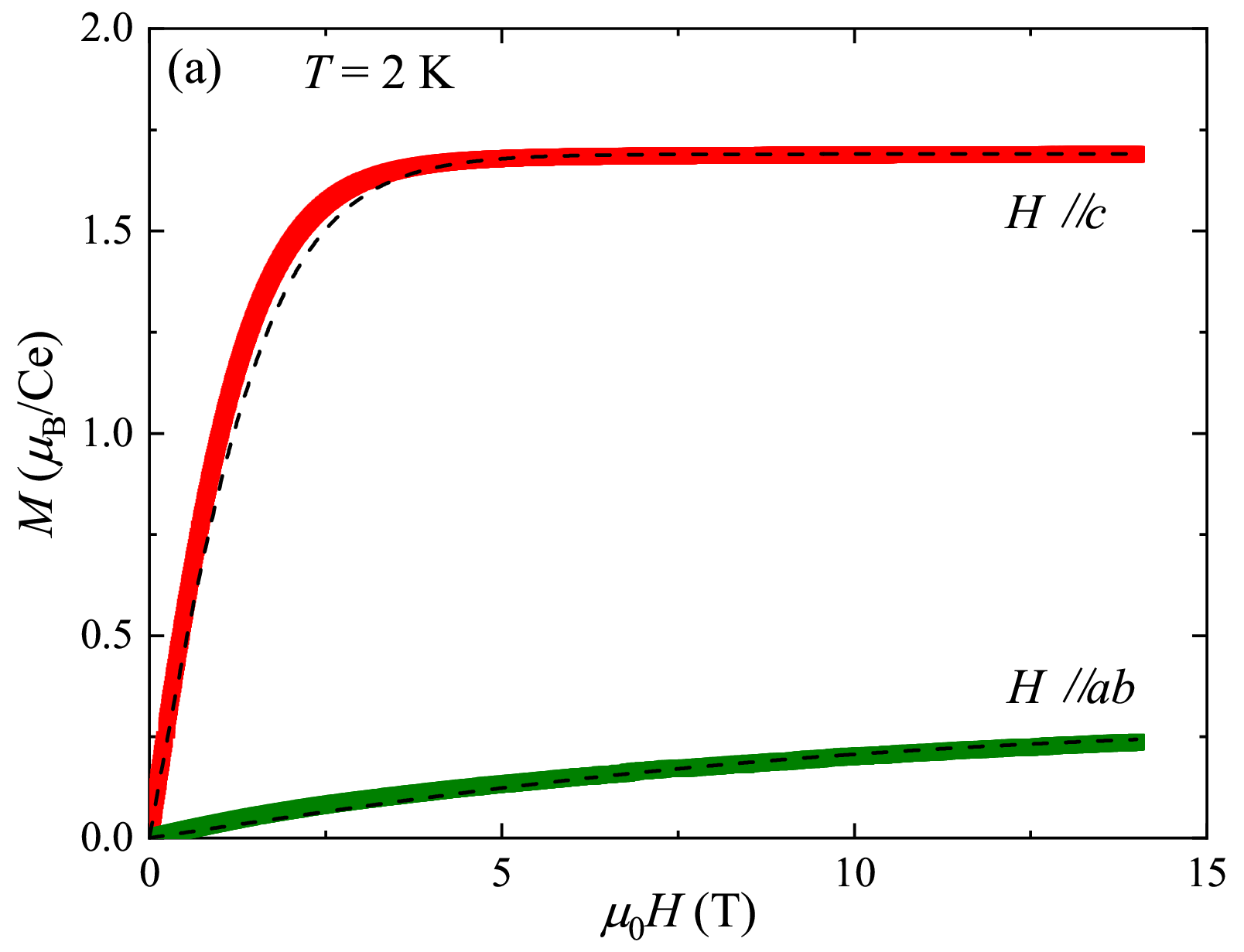}
\includegraphics[width=0.9\linewidth]{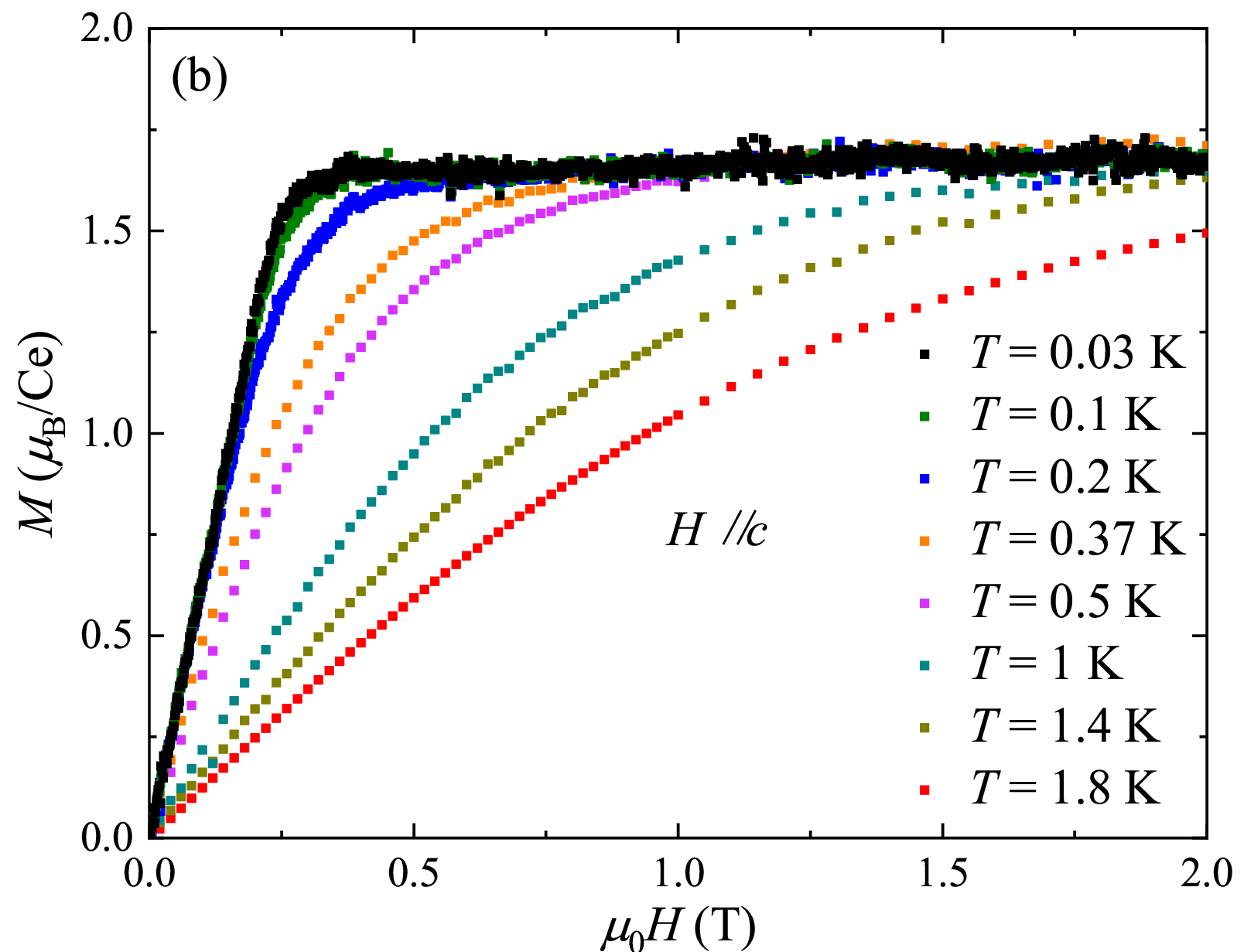}
\caption{(a) Magnetization at $T=2$\,K in CeMgAl$_{11}$O$_{19}$ along the $c$ axis and in the $ab$ plane as a function of Magnetic field. The dashed line indicates approximation of the magnetization by Brillouin functions giving the $g$ factor values: $g_c=3.4$ and  $g_{ab}\approx0.6$. (b) Magnetization of CeMgAl$_{11}$O$_{19}$ along the $c$ axis at different temperatures from $T=0.03\,$K to $T=1.8\,$K.}
    \label{MvsH} 
\end{center}
\end{figure}

Magnetization measurements at magnetic fields up to $\mu_\mathrm{0} H=14\,$T at a temperature of $T=2$\,K indicate a clear magnetic saturation along the easy magnetization axis $H\parallel c$ at $M_\mathrm{sat}=1.68\,\mu_\mathrm{B}$/Ce. It implies an effective $g$ factor of $g_c=3.36$ in a relatively good agreement with previous ESR studies in La$_{1-x}$Ce$_{x}$MgAl$_{11}$O$_{19}$ and CeMgAl$_{11}$O$_{19}$ giving $g_c=3.74$ and $g_c=3.66$, respectively~\cite{Viana1988, Gao2024}. The $g$ factor in the $ab$ plane was estimated at $g_{ab}\approx0.6$ upon approaching the magnetization by a Brillouin function. It implies a large anisotropy of the effective $g$-tensor with $g_c/g_{ab}\approx 6$. This anisotropy of the magnetization contradicts magnetization data showing a much weaker anisotropy in Ref.~\cite{Gao2024} and it agrees with data reported in Ref.~\cite{Cao2025}.

The isothermal magnetization along the $c$ axis in CeMgAl$_{11}$O$_{19}$ at different temperatures down to $T=0.03\,$K is represented in Fig.~\ref{MvsH}(b). At the lowest temperature of $T=0.03\,$K the magnetization is linear in applied magnetic field until the saturation field of $\mu_\mathrm{0} H_\mathrm{sat}\approx0.25\,$T. The absence of any magnetization plateau is consistent with the assumption of an anisotropy of the magnetic interaction $J_z<J_\perp$~\cite{Yamamoto2014, Sellmann2015}. The saturation field  is  relatively weak with a ratio of $g_c\mu_\mathrm{B}\mu_\mathrm{0} H_\mathrm{sat}/J_\perp \approx 0.9$, which is far below the theoretical value at the Heisenberg limit $g\mu_\mathrm{B}\mu_\mathrm{0} H_\mathrm{sat}/J$= 9/2~\cite{Kawamura1985, Yamamoto2014, Sellmann2015}. This result is qualitatively  consistent with a proximity of CeMgAl$_{11}$O$_{19}$ to quantum criticality towards ferromagnetism. 

\begin{figure}
\begin{center}
\includegraphics[width=0.9\linewidth]{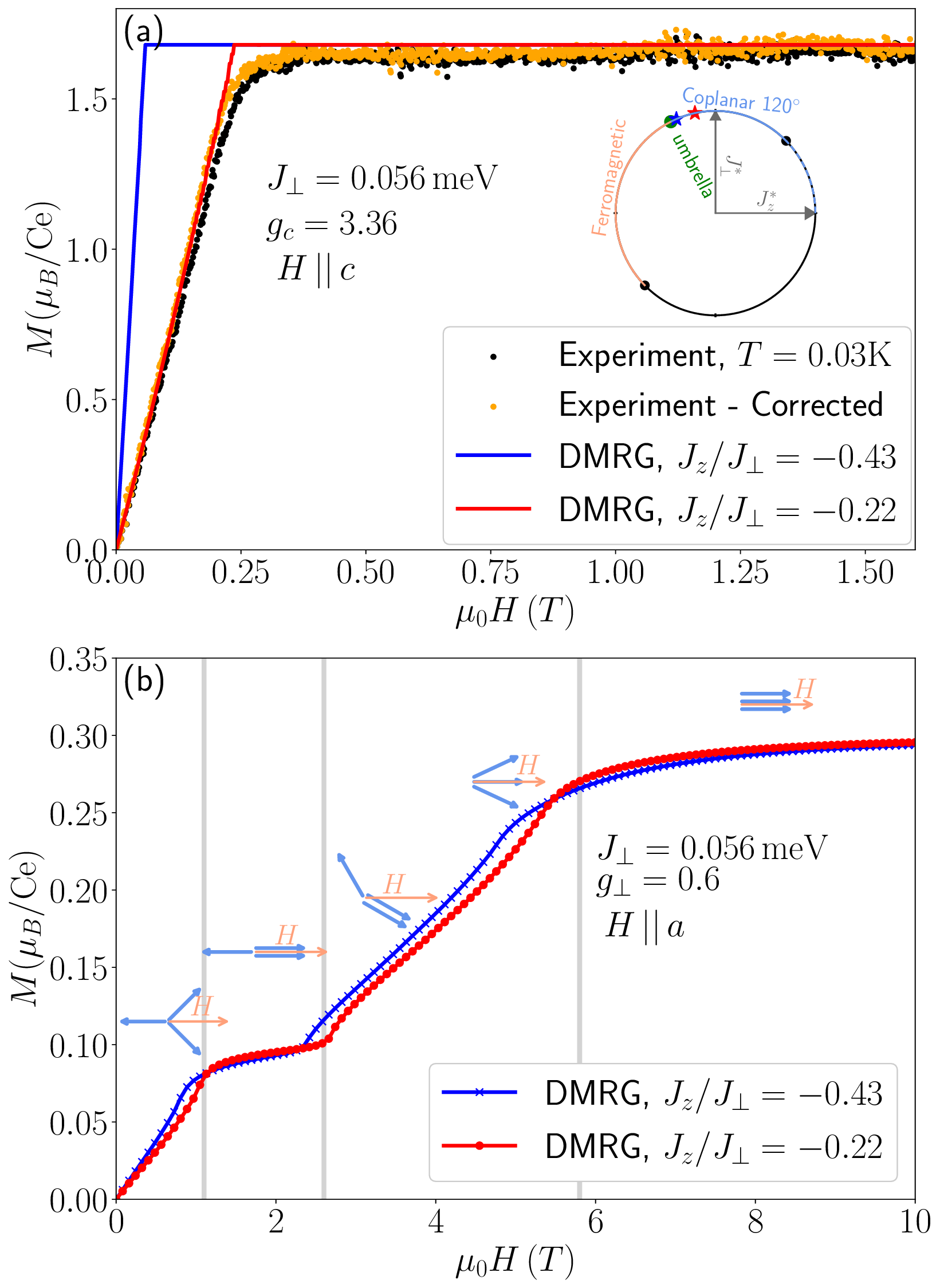}
\caption{(a) Comparison of the experimental magnetization at $T=0.03$\,K in CeMgAl$_{11}$O$_{19}$ along the $c$ axis as a function of magnetic field with the DMRG ground-state calculations. The black and orange dot corresponds to the magnetization as measured and after correction of the demagnetization effect, respectively. The red line is our best fit where $J_z/J_\perp = -0.22$ and the blue line shows results for $J_z/J_\perp = -0.43$, which was the value extracted from the best fit of INS data in Ref.~\cite{Gao2024}. The circle in the inset shows approximate ground-state phase diagram in zero magnetic field. Here $J^*_z=J_z/\sqrt{J_z^2 + J_\perp^2}$ and $J^*_\perp=J_\perp/\sqrt{J_z^2 + J_\perp^2}$. 
(b) Theoretical ground state magnetization along the $a$ axis as a function of magnetic field for $J_z/J_\perp = -0.43$ (blue symbols) and $J_z/J_\perp = -0.22$ (red symbols), predicting a 1/3 magnetization plateau. Inserted schematics show the different spin textures, which were determined from structure factor calculations (see SM)}
    \label{DMRG} 
\end{center}
\end{figure}

To quantify the proximity of CeMgAl$_{11}$O$_{19}$ to ferromagnetic quantum criticality, we analyzed magnetization using DMRG calculations based on Hamiltonian~\eqref{H} for various sets of parameters. The calculations were performed using the TeNPy library (version 1.0.4)~\cite{tenpy2024} with a lattice size of $L=12\times9$ and cylindrical boundary conditions. Details about the calculation method can be found in Supplemental Materials~\cite{sup}. To test the effect of the ratio, $J_z/J_\perp$, we kept the strongest interaction at a fixed value $J_\perp=-0.056\,$meV from Ref.~\cite{Gao2024} and we performed calculation for various values of $J_z$. The computational data at $T=0$ were compared with the experimental data measured at $T=0.03\,$K in Fig.~\ref{DMRG}(a). To check whether the demagnetization effect could influence the fitting by DMRG, the demagnetization effect was estimated using the approximation proposed in Ref.~\cite{Aharoni1998}. It implies a minor correction that was taken into account in analyzing experimental results. 

Our best fit value, $J_z/J_\perp = -0.22$ confirms the presence of strongly anisotropic magnetic interactions which ferromagnetically couple the spin component $S^z$. However, it places CeMgAl$_{11}$O$_{19}$ further away from the quantum critical point at $J_z/J_\perp = -0.5$ than fits of INS data suggesting $J_z/J_\perp = -0.43$~\cite{Gao2024}. Furthermore, DMRG calculations predict the formation of a 120$^\circ$ for $J_z/J_\perp = -0.22$ and its evolution into an umbrella state under magnetic field as shown in Supplemental Material with structure factor calculations~\cite{sup}. Therefore, our experimental results contradict the proposal of a quantum spin liquid ground state induced by a close proximity to the exactly solvable point $J_z/J_\perp = -0.5$.

\begin{figure}[t]
\includegraphics[width=0.87\linewidth]{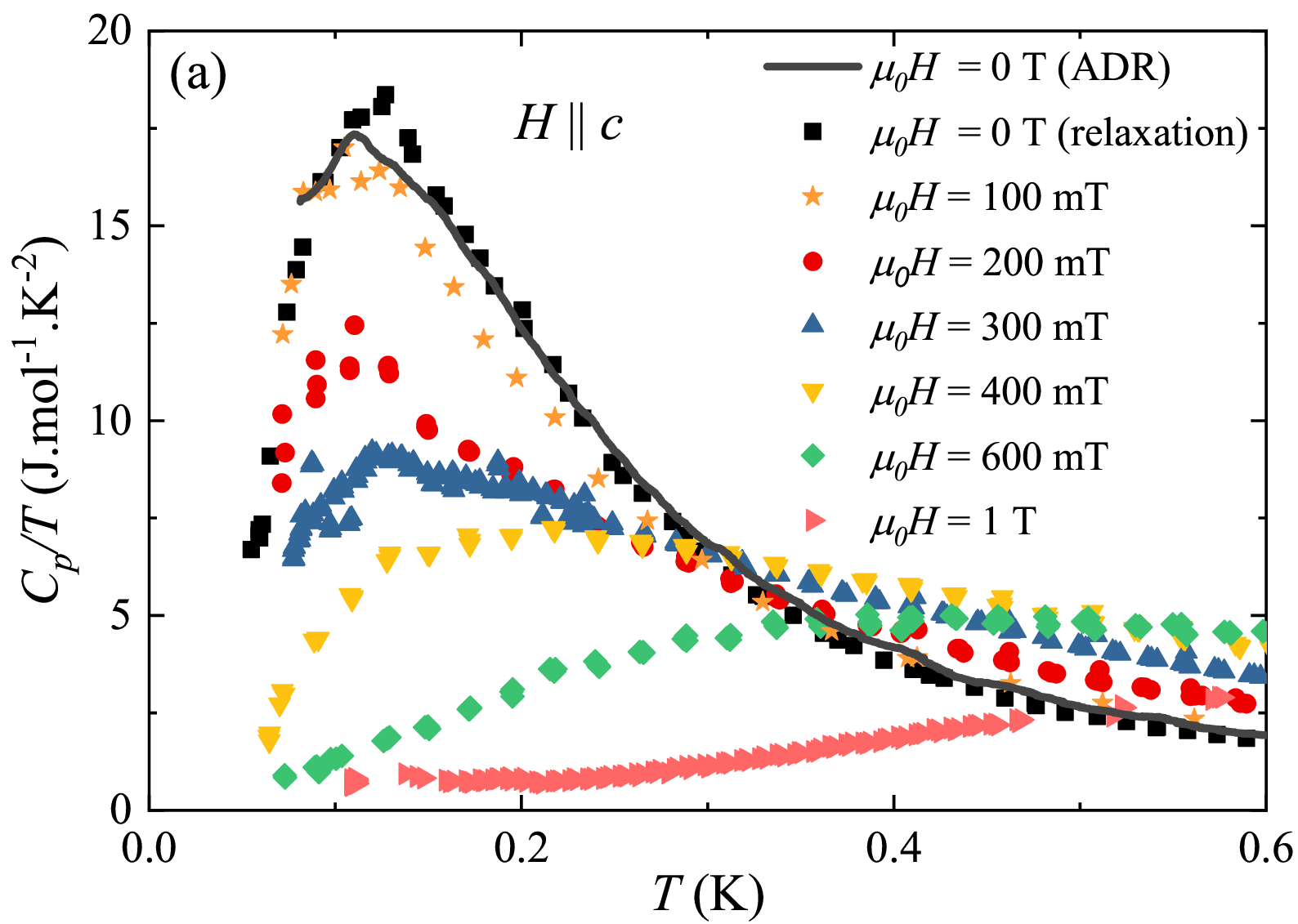}
\includegraphics[width=0.87\linewidth]{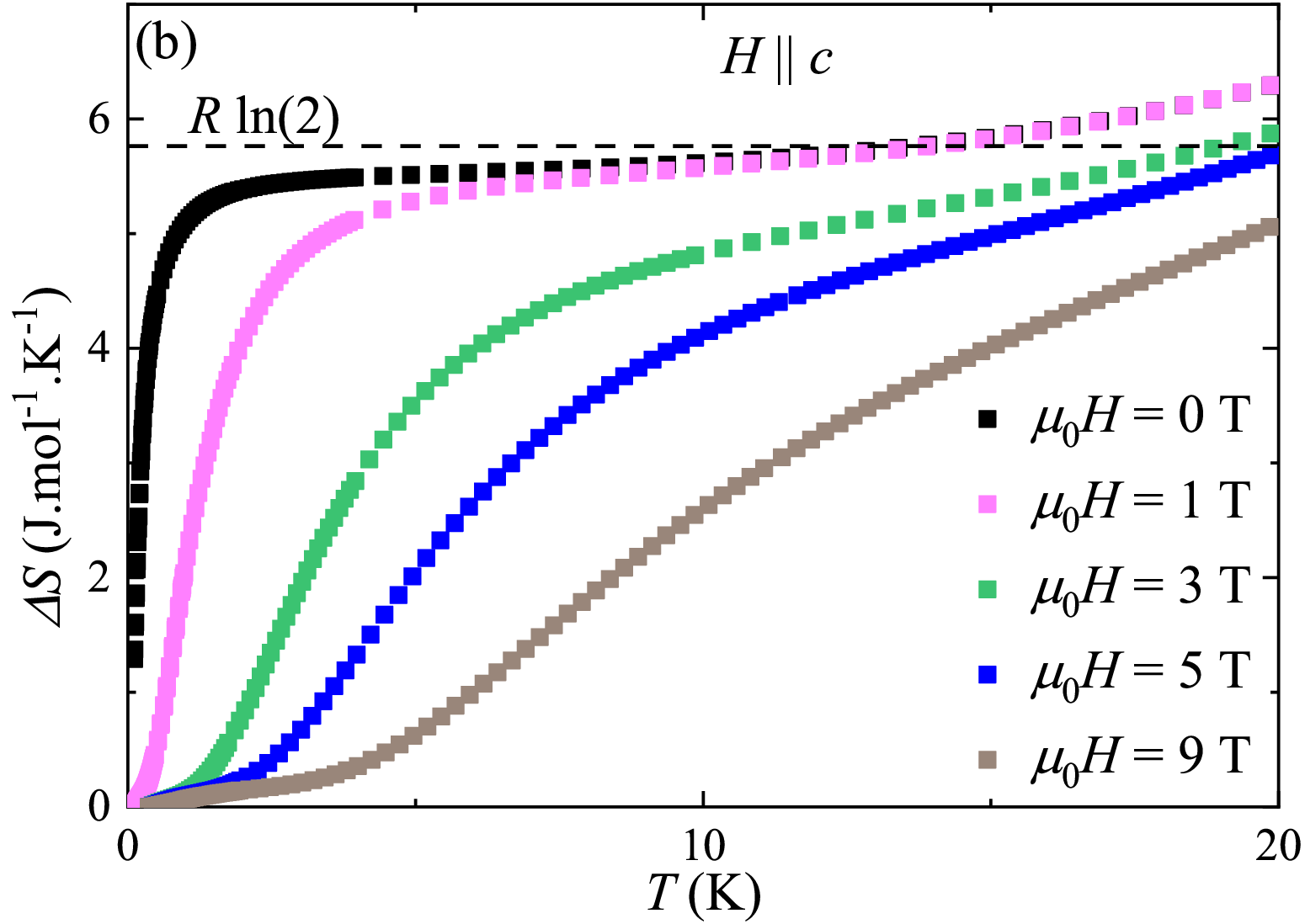}
\caption{(a) $C_p/T$ in CeMgAl$_{11}$O$_{19}$ below $T=$0.6\,K as a function of temperature and magnetic field applied along the $c$ axis. The zero-field specific heat was measured both with the relaxation method and adiabatic demagnetization refrigeration. (b) Variation of entropy $\Delta S= S - S(T=0.4\,$K$, \mu_\mathrm{0}H=9\,$T) as a function of temperature and magnetic field. The horizontal dashed line indicates the expected value of $R\mathrm{ln}(2)$}
\label{Cp}
\end{figure}

The zero-temperature magnetization in the $ab$ plane computed by DMRG is represented in Fig.~\ref{DMRG}(b). These calculations predict a magnetization plateau with $M/M_{\text{sat}}\approx 1/3$ for $\mu_\mathrm{0}H\approx 1.8$\,T, reflecting the formation of the up-up-down order~\cite{Kawamura1985, Yamamoto2014, Sellmann2015}. Unfortunately, these predictions could not be tested, due to the low effective $g-$factor in the $ab$ plane preventing any measurement of the magnetization in the $ab$ plane below $T=1.8\,$K.

The specific heat divided by temperature $C_p/T$ in CeMgAl$_{11}$O$_{19}$ as a function of temperature below $T=0.6\,$K is represented in Fig.~\ref{Cp}(b). It does not show any peak which would indicate long-range magnetic ordering. On the contrary, $C_p/T$ in absence of external magnetic field harbors a broad maximum around $T\approx0.12\,$K. This maximum broadens under the application of magnetic field up to the saturation field of $\mu_\mathrm{0}H\approx0.25\,$T excluding the formation of magnetic field-induced long-range magnetic order. Under magnetic fields higher than the saturation field $\mu_\mathrm{0} H_\mathrm{sat} = 0.25\,$T, the maximum of $C_p/T$ shifts to higher temperature with increasing field as commonly observed in ferromagnets~\cite{Bastien2024}. 

The entropy variation with temperature and magnetic field was obtained based on the integration of magnetization and specific heat as explained in Ref.~\cite{Bastien2020} and it is represented in Fig.~\ref{Cp}(b). The difference of entropy between magnetic saturation at $T=0.4\,$K and $\mu_\mathrm{0} H > 3\,$T and the paramagnetic regime at $5\,$K$<T< 15\,$K and $\mu_\mathrm{0} H=0$ approaches $\Delta S=R\mathrm{ln}(2)$. This result further confirms the formation of effective spins $S_\mathrm{eff}=1/2$. The strong dependence of the entropy on magnetic field implies the possibility to cool CeMgAl$_{11}$O$_{19}$ samples using adiabatic demagnetization of the sample itself. We demonstrated this possibility in the same setup used in Refs.~\cite{Tokiwa2021, Treu2024}. Demagnetization from $\mu_\mathrm{0}H=5\,$T at $T=2\,$K yields a final temperature of
$T=0.09\,$K. The subsequent warm-up curve was used to calculate $C_p(T)$, displayed as line in Fig.~\ref{Cp}(a).


There is an apparent contradiction between the DMRG study predicting  the 120$^\circ$ order for $J_z/J_\perp = -0.22$ and the absence of magnetic ordering in CeMgAl$_{11}$O$_{19}$ down to 0.03\,K. 
This contradiction comes most likely from structural disorder that affects the magnetic phase diagram of the TLAF~\cite{Maryasin2013, Dey2020, Li2017, Kimchi2018, Zhu2017,Cairns2024} beyond our DMRG calculations. The most studied disordered-induced quantum magnetic state is the random singlet (RS) state proposed in YbMgGaO$_4$~\cite{Li2017,Kimchi2018, Li2020}, however the occurrence of this phase in CeMgAl$_{11}$O$_{19}$ is unlikely since power law divergence of $\chi$ and $C_p/T$ towards $T=0$ were predicted for a RS state~\cite{Kimchi2018, Kimchi2018a} in contradiction with our data. Alternatively, we could propose the formation short-range 120$^\circ$  magnetic order in CeMgAl$_{11}$O$_{19}$ as previously proposed in YbMgGaO$_4$~\cite{Zhu2017} and in NaYbSe$_2$~\cite{Cairns2024}. 

To conclude, our single crystal study of the $S_\mathrm{eff}=1/2$ effective spin TLAF CeMgAl$_{11}$O$_{19}$ confirm the absence of magnetic ordering down to $T=0.03\,$K. The combination of magnetization measurement down to $T=0.03\,$K and DMRG calculation confirms the proposal of strongly anisotropic magnetic interactions antiferromagnetically coupling in-plane spin components and ferromagnetically coupling the out of plane spin component. However it excludes the close proximity of CeMgAl$_{11}$O$_{19}$ to the ferromagnetic quantum critical point, where a spin liquid was analytically predicted. Finally, we argue that the disordered magnetic ground state of CeMgAl$_{11}$O$_{19}$ arises from the combination of magnetic frustration, structural disorder and proximity to quantum criticality and the question whether this magnetic state can be classified as a quantum spin liquid remains open for future debate. 

\begin{acknowledgments}
We acknowledge funding from the Charles University in Prague within the Primus research program with grant number PRIMUS/22/SCI/016. The work was also supported by the Ministry of Education, Youth and Sports of the Czech Republic through programs e-INFRA CZ (ID:90254), and INTER-EXCELLENCE II INTER-ACTION (LUABA24056). M.\v{Z}. acknowledges support from the Czech Science Foundation via Project No.22-22419S. The work in Augsburg was funded by the Deutsche Forschungsgemeinschaft (DFG, German Research Foundation), Grants No. 514162746 (GE 1640/11-1) and No. 492547816 (TRR 360) as well as by the Bavarian-Czech University Agency, project number BTHA-JC-2024-15. Crystal growth, structural analysis and magnetic properties measurements were carried out in the MGML (http://mgml.eu/), which is supported within the program of Czech Research Infrastructures (project no. LM2023065). We acknowledge \v{S}\'arka Matou\v{s}kov\'a and Jan Rohovec from the Institute of Geology of the Czech Academy of Sciences (RVO 67985831) for her help for the determination of crystal composition. 

\end{acknowledgments}

\bibliography{CMAO-PRL}

\clearpage                   
\onecolumngrid               
\section*{Supplemental Material}
\beginsupplement


\section{Composition analysis by x-ray fluorescence (XRF)  spectroscopy}

\begin{figure}[h!] 
\begin{center} 
\includegraphics[width=0.8\linewidth]{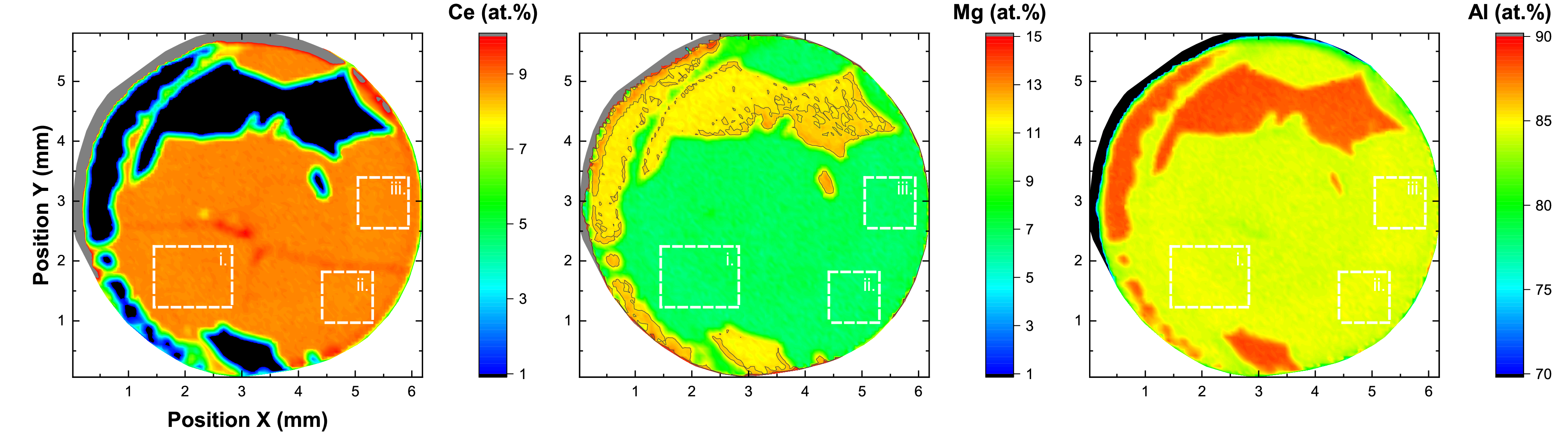} 
\caption{Results of X-ray fluoresence composition mapping of a disc cut from the grown ingot. The dashed white boxes indicate the regions used for magnetometry.} 
\label{fig:XRF_map} 
\end{center} 
\end{figure}

The floating-zone grown oligocrystal ingot was sliced by two cuts transverse to growth direction using a diamond band saw. The cross-section of the resulting disc-shaped sample was investigated by X-ray fluorescence spectroscopy to identify the grains of correct composition. The composition analysis was performed on an EDAX Orbis spectrometer using a Rh anode source, and polycapillary focusing optics. Spectral processing used standardless fundamental principles for composition extraction, and uncertainties are estimated to be in the range of 1.0 at.\%.
Full mapping of the disc, shown in Fig.~\ref{fig:XRF_map} , resulted in the observation of regions with a Ce content far below the nominal value. These were confirmed by powder X-ray diffraction to be magnesium aluminate spinel, MgAl$_2$O$_4$. Compositionally homogeneous single-grain regions were identified, and extracted by precision wire-saw cutting. The white boxes indicated in Fig.~\ref{fig:XRF_map} shows the isolated region investigated within this manuscript, with an average (mode) determined cation composition of 8.8 : 6.7 : 84.5 at.\%, for Ce : Mg : Al, respectively. This is within error of the expected cationic  ratio (Ce : Mg : Al, 1 : 1 : 11) of 7.7 : 7.7 : 84.6 at.\%, specifically considering the large difference in fluoresence yield when comparing very light (Mg, Al) and very heavy (Ce) elements. 

\section{Detailed results from single crystal x-ray diffraction}

A single-crystal X-ray diffraction experiment was performed
at 120\,K on an Xcalibur, Atlas, Gemini ultra four-circle diffractometer, with graphite-monochromated Mo K$\alpha$ ($\lambda$ = 0.71073\,$\mathring{\mathrm{A}}$) radiation, equipped with an Atlas S2 CCD detector. Diffraction data were integrated using CrysAlis Pro~\cite{crysalis} with an empirical absorption correction using spherical harmonics~\cite{Clark1995} combined with an analytical numeric absorption correction based on Gaussian integration on a multifaceted crystal model implemented in the SCALE3 ABSPACK scaling algorithm. The structure was solved by charge flipping using the program Superflip~\cite{Palatinus2007} and refined by full-matrix least squares on $F^2$ in Jana2020~\cite{Petricek2023}. 


\begin{table}[h]
    \caption{Single crystal structure solution and refinement information}
	\begin{tabular}{cccccccccccccccccccccccccc}
		\hline
		
		\multicolumn{5}{c}{\begin{tabular}[c]{@{}c@{}}Ce$_{0.8954}$Mg$_{0.378}$Al$_{11.622}$O$_{19}$\\ Crystal system \& Space   group: \\    hexagonal, $P6_3/mmc$ (\#194, setting 1)\end{tabular}}                                                 & \multicolumn{5}{c}{$a$ = 5.5789(5) $\mathring{\mathrm{A}}$}                                                         & \multicolumn{5}{c}{$c$ = 21.9124(16)
		 $\mathring{\mathrm{A}}$}                                                                                                          & \multicolumn{5}{c}{$V$ = 590.64(8)
		  $\mathring{\mathrm{A}}^3$}                                                                                                               & \multicolumn{5}{c}{$Z$ = 2}                                                                                                                           \\
		
		\hline
		
		\multicolumn{4}{c}{\begin{tabular}[c]{@{}c@{}}Radiation:   \\    X-ray tube: Mo K$\alpha$ \\    ($\lambda$ = 0.71073 $\mathring{\mathrm{A}}$)\end{tabular}} & \multicolumn{3}{c}{\begin{tabular}[c]{@{}c@{}}Reflections   \\ collected/unique/used: \\ 8295/357/305\\
    Parameters: 44\end{tabular}} & \multicolumn{3}{c}{\begin{tabular}[c]{@{}c@{}}Final $R$   indices: \\    $R_{obs}$ = 2.80 \% \\    $wR_2$ = 7.04 \% \\ GOF = 1.3402\end{tabular}} & \multicolumn{7}{c}{\begin{tabular}[c]{@{}c@{}}Maximum   difference:\\  Peaks \& Holes ($e\mathring{\mathrm{A}}^{-3}$) \\        1.67   \&   $-$1.37 \end{tabular}} & \multicolumn{4}{c}{\begin{tabular}[c]{@{}c@{}}Density: \\    $\rho$ = 4.2297 g.cm$^{-3}$\end{tabular}} & \multicolumn{4}{c}{\begin{tabular}[c]{@{}c@{}}  Absorption \\ coefficient: \\    $\mu$ = 4.506 mm$^{-1}$\end{tabular}} \\
		
		\hline
		
		\multirow{2}{*}{$T$ (K)}             & \multirow{2}{*}{Atom}            & \multirow{2}{*}{Site}            & \multicolumn{3}{c}{\multirow{2}{*}{$x$}}                                                           & \multicolumn{2}{c}{\multirow{2}{*}{$y$}}                                                 & \multirow{2}{*}{$z$}                                        & \multicolumn{2}{c}{\multirow{2}{*}{Occ.}}                   & \multicolumn{15}{c}{$U_{\mathrm{ij}}$ ($\times$10$^{-4}  \mathring{\mathrm{A}}^2$)}                                                                                                                                                                                                                                             \\ \cline{12-25}
		&                                  &                                  & \multicolumn{3}{c}{}                                                                             & \multicolumn{2}{c}{}                                                                   &                                                           & \multicolumn{2}{c}{}                                        & \multicolumn{2}{c}{$U_{11}$}                    & \multicolumn{2}{c}{$U_{22}$}                 & \multicolumn{2}{c}{$U_{33}$}              & \multicolumn{3}{c}{$U_{12}$}                          & \multicolumn{2}{c}{$U_{13}$}       & \multicolumn{2}{c}{$U_{23}$}       & \multicolumn{1}{c}{$U_{\mathrm{eq}}$}            &               \\
		\hline
		
		\multirow{12}{*}{120 K}             & Ce                               & 2d                               & \multicolumn{3}{c}{$\frac{1}{3}$}                                                                          & \multicolumn{2}{c}{$\frac{2}{3}$}                                                                  & $\frac{3}{4}$                                                      & \multicolumn{2}{c}{0.895(4)}                                       & 89(3)                  & \multicolumn{3}{c}{$U_{11}$}                                      & \multicolumn{3}{c}{36(5)}                           & \multicolumn{2}{c}{$\frac{1}{2}U_{11}$}    & \multicolumn{2}{c}{0}         & \multicolumn{2}{c}{0}         & 72(2)              &               \\
		
		& Al(1)                            & 12k                              & \multicolumn{3}{c}{0.16763(12)}                                                                   & \multicolumn{2}{c}{0.3353(2)}                                                         & 0.60823(5)                                                & \multicolumn{2}{c}{1}                                       & 37(6)                  & \multicolumn{3}{c}{35(5)}                                      & \multicolumn{3}{c}{40(7)}                           & \multicolumn{2}{c}{$\frac{1}{2} U_{22}$}      & \multicolumn{2}{c}{2(5)}      & \multicolumn{2}{c}{$2U_{13}$}      & 37(5)              &               \\

		& Al(2)                            & 4f                               & \multicolumn{3}{c}{$\frac{1}{3}$}                                                                            & \multicolumn{2}{c}{$\frac{2}{3}$}                                                                  & 0.19004(9)                                                & \multicolumn{2}{c}{1}                                       & 34(7)                  & \multicolumn{3}{c}{$U_{11}$}                                      & \multicolumn{3}{c}{37(11)}                           & \multicolumn{2}{c}{$\frac{1}{2}U_{11}$}    & \multicolumn{2}{c}{0}         & \multicolumn{2}{c}{0}         & 35(6)              &               \\
		
		& Al(3)                            & 2a                               & \multicolumn{3}{c}{0}                                                                            & \multicolumn{2}{c}{0}                                                                  & 0                                                         & \multicolumn{2}{c}{1}                                       & 37(9)                  & \multicolumn{3}{c}{$U_{11}$}                                      & \multicolumn{3}{c}{26(17)}                            & \multicolumn{2}{c}{$\frac{1}{2}U_{11}$}    & \multicolumn{2}{c}{0}         & \multicolumn{2}{c}{0}         & 33(8)              &               \\
		
		& Al(4)                            & 4f                               & \multicolumn{3}{c}{$\frac{1}{3}$}                                                                            & \multicolumn{2}{c}{$\frac{2}{3}$}                                                                  & 0.02749(10)                                                & \multicolumn{2}{c}{0.7(2)}                                       & 38(16)                  & \multicolumn{3}{c}{$U_{11}$}                                      & \multicolumn{3}{c}{56(18)}                           & \multicolumn{2}{c}{$\frac{1}{2}U_{11}$}    & \multicolumn{2}{c}{0}         & \multicolumn{2}{c}{0}         & 44(12)              &               \\
		& Mg(1)                            & 4f                               & \multicolumn{3}{c}{$\frac{1}{3}$}                                                                            & \multicolumn{2}{c}{$\frac{2}{3}$}                                                                  & 0.02749(10)                                                & \multicolumn{2}{c}{0.3(2)}                                       & 38(10)                  & \multicolumn{3}{c}{$U_{11}$}                                      & \multicolumn{3}{c}{56(18)}                           & \multicolumn{2}{c}{$\frac{1}{2}U_{11}$}    & \multicolumn{2}{c}{0}         & \multicolumn{2}{c}{0}         & 44(12)             &               \\
		& Al(5)                            & 4e                               & \multicolumn{3}{c}{0}                                                                            & \multicolumn{2}{c}{0}                                                                  & 0.2410(6)                                               & \multicolumn{2}{c}{$\frac{1}{2}$}                                       & 46(10)                  & \multicolumn{3}{c}{$U_{11}$}                                      & \multicolumn{3}{c}{70(90)}                           & \multicolumn{2}{c}{$\frac{1}{2}U_{11}$}    & \multicolumn{2}{c}{0}         & \multicolumn{2}{c}{0}         & 60(30)              &               \\
		& O(1)                             & 6h                               & \multicolumn{3}{c}{0.1817(4)}                                                                    & \multicolumn{2}{c}{0.3634(18)}                                                          & $\frac{1}{4}$                                                         & \multicolumn{2}{c}{1}                                       & 100(16)                  & \multicolumn{3}{c}{70(20)}                                   & \multicolumn{3}{c}{67(18)}                          & \multicolumn{2}{c}{$\frac{1}{2}U_{11}$}    & \multicolumn{2}{c}{0}         & \multicolumn{2}{c}{0}         & 81(15)              &               \\
		
		& O(2)                             & 12k                              & \multicolumn{3}{c}{0.5049(3)}                                                                    & \multicolumn{2}{c}{0.0099(5)}                                                        & 0.15098(13)                                                & \multicolumn{2}{c}{1}                                       & 38(9)                 & \multicolumn{3}{c}{$U_{11}$}                                    & \multicolumn{3}{c}{46(16)}                           & \multicolumn{2}{c}{10(11)}    & \multicolumn{2}{c}{-5(5)}      & \multicolumn{2}{c}{$-U_{13}$}      & 44(9)             &               \\
		
		& O(3)                             & 4e                               & \multicolumn{3}{c}{0}                                                                            & \multicolumn{2}{c}{0}                                                                  & 0.1509(2)                                               & \multicolumn{2}{c}{1}                                       & 44(14)                  & \multicolumn{3}{c}{$U_{11}$}                                      & \multicolumn{3}{c}{60(30)}                          & \multicolumn{2}{c}{$\frac{1}{2}U_{11}$}    & \multicolumn{2}{c}{0}         & \multicolumn{2}{c}{0}         & 50(13)              &               \\
		
		& O(4)                             & 12k                              & \multicolumn{3}{c}{0.1525(2)}                                                                  & \multicolumn{2}{c}{0.3051(5)}                                                          & 0.05316(14)                                                & \multicolumn{2}{c}{1}                                       & 50(10)                  & \multicolumn{3}{c}{$U_{11}$}                                    & \multicolumn{3}{c}{69(15)}                           & \multicolumn{2}{c}{20(11)}    & \multicolumn{2}{c}{10(6)}    & \multicolumn{2}{c}{$-U_{13}$}      & 59(10)              &               \\

		& O(5)                             & 4f                               & \multicolumn{3}{c}{$\frac{1}{3}$}                                                                            & \multicolumn{2}{c}{$\frac{2}{3}$}                                                                  & 0.5577(2)                                               & \multicolumn{2}{c}{1}                                       & 29(14)                 & \multicolumn{3}{c}{$U_{11}$}                                      & \multicolumn{3}{c}{80(30)}                          & \multicolumn{2}{c}{$\frac{1}{2}U_{11}$}    & \multicolumn{2}{c}{0}         & \multicolumn{2}{c}{0}         & 47(13)              &              
	\end{tabular}
\end{table}

\begin{figure}[h!] 
\begin{center} 
\includegraphics[width=0.8\linewidth]{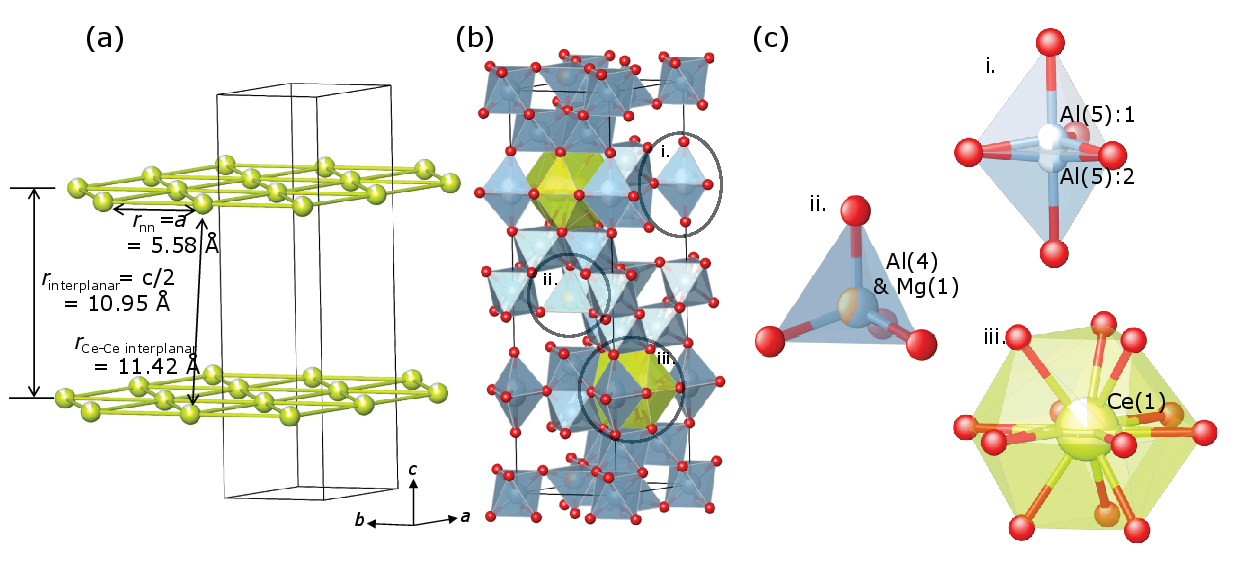} 
\caption{Structural representations, created using Vesta ~\cite{Momma2008}. (a) Highlighting the well separated triangular lattices of Ce$^{3+}$ ions, with both interplanar, and nearest neightbour Ce-Ce ion distances. Note that due to offset stacking, the interplanar Ce-Ce distance is larger than $c/2$. (b) Shows the full structure, and (c) highlights the three sources of inherent disorder, described above.} 
\label{fig:Structure} 
\end{center} 
\end{figure}

The structure is shown in Fig.~\ref{fig:Structure}.
It is important to note that there are three sources of structural disorder considered in this refinement: Ce deficiency (Fig.~\ref{fig:Structure} ciii), random distribution of Mg on Al(4) site (Fig.~\ref{fig:Structure} cii) and the splitting of Al(5) between two sites on both sides of the horizontal mirror plane (Fig.~\ref{fig:Structure} ci). The refinement reported above does not include the off-centering of the Ce ion in its oxygen cage, which was previously observed in LnMgAl$_{11}$O$_{19}$ (Ln = La-Pr)~\cite{Abrahams1987, Cao2024, Cao2025, Kumar2025}. However, another refinement was attempted, taking this possible disorder effect into account, and it resulted in the off-centering of only 0.021(9) Ce per formula unit, with only a minor improvement of the fit. Therefore, we prefer to give the refinement results without considering the possibility of off-centering of the Ce ion.

\section{Detailed method for the DFT calculations}

Band structure calculations were first performed using the projector augmented wave (PAW) method in \emph{MedeA-VASP} program package~\cite{Kresse1996} in order to optimize the unit cell parameters and atomic positions within the $P6_3/mmc$ space group.

The optimized crystal structure was then used to obtain the self-consistent
electron density and potential using an all-electron full potential code as implemented in the WIEN2k package~\cite{Blaha2001}. The augmented plane waves + local orbital basis set ($RK_{max}$ parameter 7.0) and the generalized
gradient approximation for the exchange correlation functional~\cite{Perdew1996} were applied. The atomic 
(muffin tin) radii of 2.0, 1.7, 1.8 and 1.6\,a.u. were selected for Ce and Mg, Al and O, respectively.
The first Brillouin zone sampling $9\times 9 \times 2$ and tetrahedron integration scheme were applied. 

In the first step, the calculation was performed as non-spin polarized with Ce-$4f$ states included in the core. In the second step, the eigenvalue problem was solved in a single run using \emph{WIEN2k} (\emph{LAPW1} program) with Ce-$4f$ states considered as valence states. Next the wave functions in plane-wave representation were transformed into Wannier functions by 
means of \emph{Wien2wannier} \cite{Kunes2010} and \emph{Wannier90} \cite{Mostofi2008} programs. The Ce-$4f$ states were considered as valence states and O-$2s, 2p$ states were shifted downwards by an energy parameter ($\Delta = 20$\,eV and 8\,eV, respectively)
by applying an orbital dependent potential. This empirical parameter controls the Ce-$4f$ 
\textendash\ O-$2p, 2s$ hybridization.
Next the wave functions in plane-wave representation were transformed into
Wannier functions by means of Wien2wannier~\cite{Kunes2010} and Wannier90~\cite{Mostofi2008} programs.

For the extraction of the magnetic susceptibility and the magnetic moment, the Zeeman terms were added
to the local Hamiltonian and applied, within the first and second order perturbation theory, on the $|M_J\rangle$ basis 
states of the $J=5/2$ multiplet. The results provide the multiplet splitting by the crystal 
and magnetic fields, as well as the magnetic susceptibility calculated using van Vleck formula
along the principal $z-$axis ($C_3$) and perpendicular to it.

\section{Calculation of the temperature dependence of the effective moment by DFT and comparison with experimental results}

Adding the Zeeman term to the crystal field Hamiltonian yields once again a diagonal 
matrix for the principal $z$-axis with linear van Vleck terms $M_J g_J \mu_\mathrm{B} H_z$, while
the $\hat{J}_x = 1/2(\hat{J}_{+} +\hat{J}_{-})$ operator results in two diagonal terms 
$\pm 3/2 g_J \mu_\mathrm{B} H_x$ for $| M_J = \pm 1/2 \rangle $ and off-diagonal terms 
$\langle \pm 5/2 | \hat{J}_x | \pm 3/2 \rangle = \sqrt{5}/2$, $\langle -3/2 | \hat{J}_x | 
\psi_{\pm} \rangle = 1$ and $\langle 3/2 | \hat{J}_x | \psi_{\pm} \rangle = \pm 1$ (all in 
$g_J \mu_\mathrm{B} H_x$ units), where $\psi_{+} = (|1/2\rangle + |-1/2\rangle)/\sqrt{2}$ and 
$\psi_{-} = (-|1/2\rangle + |-1/2\rangle)/\sqrt{2}$. Application of the first and second order
perturbation theory on the magnetic Hamiltonian yields two linear van Vleck terms, $\pm 3/2 
g_J \mu_\mathrm{B} H_x$ for $\psi_{\pm}$ and three quadratic terms (in $g_J^2 \mu_\mathrm{B}^2 H_x^2$ units)

\begin{equation*}
 W_{\pm5/2}^{(2)} = \frac{5}{4\Delta_2},\; 
 W_{\pm3/2}^{(2)} = -\frac{5}{4\Delta_2} - \frac{2}{\Delta_1},\;
 W_{\psi_{\pm}}^{(2)} = \frac{2}{\Delta_1}
\end{equation*}

\noindent where the energy parameters $\Delta_1 = E_{\pm1/2} - E_{\pm3/2} = 13.6$\,meV and 
$\Delta_2 = E_{\pm5/2} - E_{\pm3/2} = -14.1$\,meV represent the differences in zeroth order 
crystal field Hamiltonian eigenvalues.

\begin{figure}[t]
\begin{minipage}{0.49\linewidth}
\begin{center}
\includegraphics[width=1\linewidth]{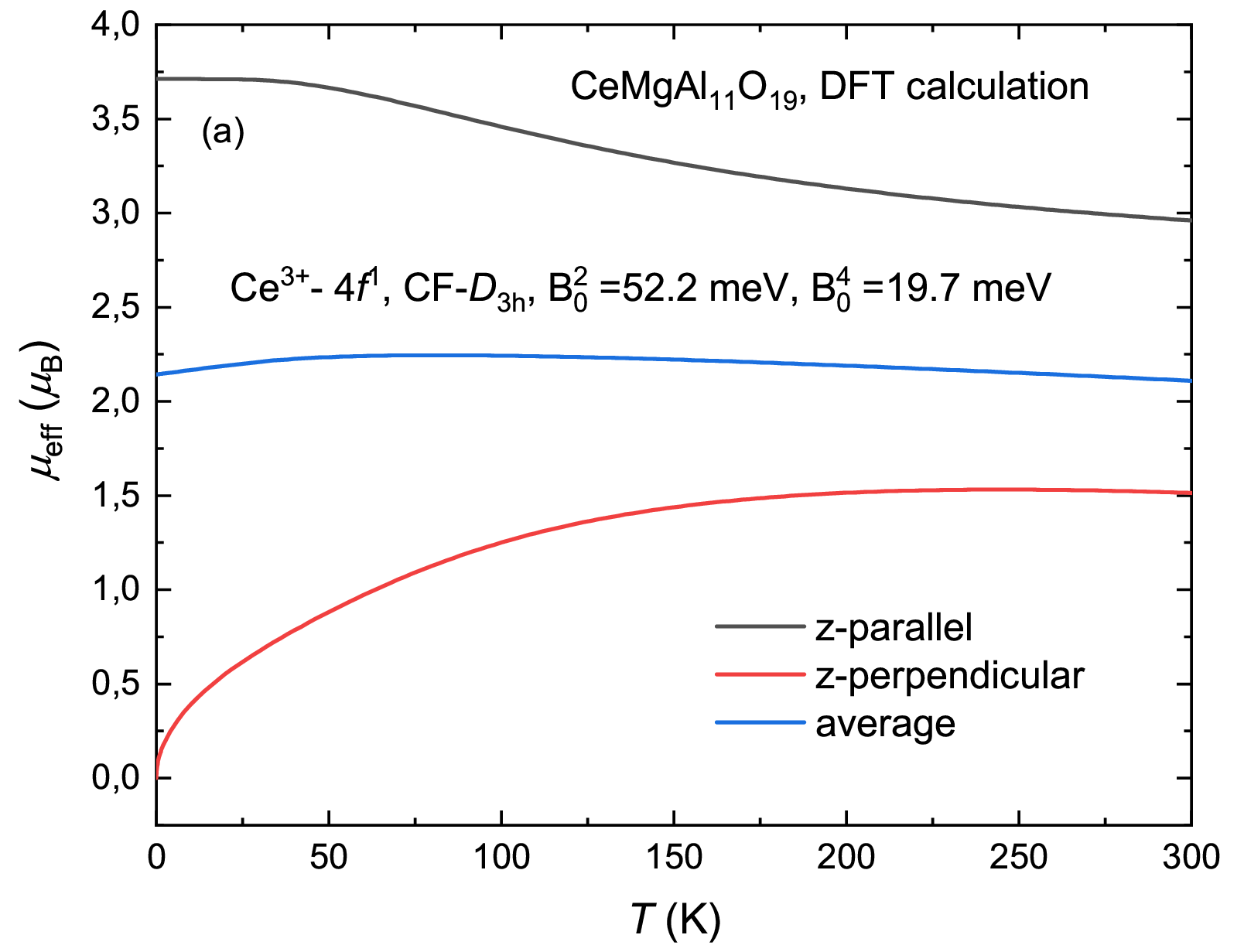}
\end{center}
\end{minipage}
\hfill
\begin{minipage}{0.49\linewidth}
\begin{center}
\includegraphics[width=0.98\linewidth]{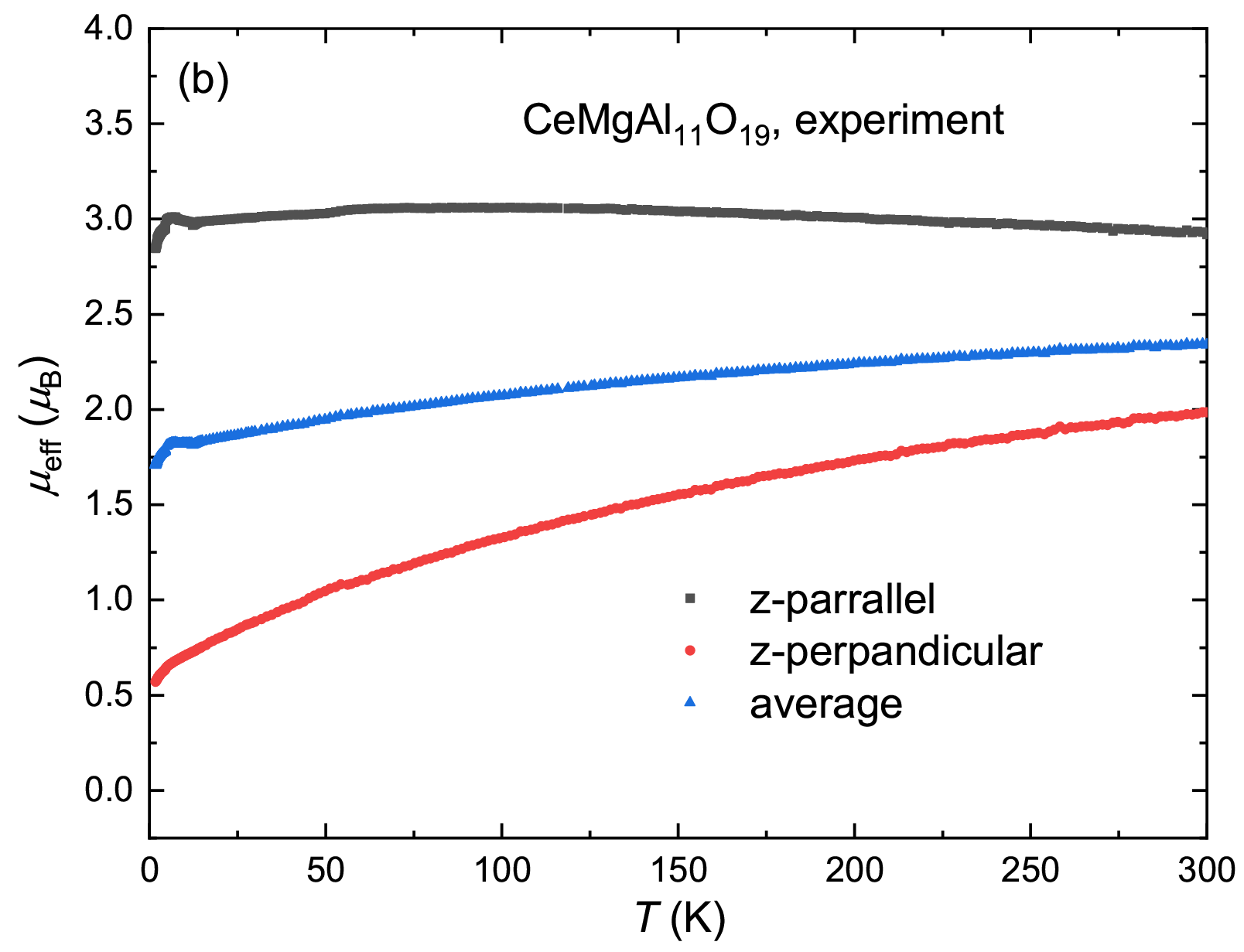}
\end{center}
\end{minipage}
  \caption{(a) Effective magnetic moment calculated using van Vleck equation for Ce$^{3+}$ ground state
  multiplet split by crystal and magnetic field applied along the principal $z$-axis and $x$-axis. (b) Experimental value of the effective moment.}
  \label{FigMueff}
\end{figure}

The effective magnetic moment shown in Fig.~\ref{FigMueff}(a) for the magnetic field parallel 
and perpendicular to the $C_3(z)$ axis was calculated by substituting the parameters given 
above into van Vleck equation modified to be consistent with Curie's law, i.e. with $\mu_\mathrm{eff}^2$
multiplied by the factor 3 (to cancel with 3 in the denominator of the Curie constant)

\begin{equation} \label{mu-par}
 \mu_{\mathrm{eff},z}^2 = 3 g_J^2 \frac{\tfrac{1}{4}\exp(-\tfrac{\Delta_1}{kT}) + \tfrac{9}{4} +
 \tfrac{25}{4}\exp(-\tfrac{\Delta_2}{kT})}
 {\exp(-\tfrac{\Delta_1}{kT}) + 1 + \exp(-\tfrac{\Delta_2}{kT})} \;,
\end{equation}

\begin{equation} \label{mu-per}
 \mu_{\mathrm{eff},x}^2 = 3 g_J^2 \frac{(\tfrac{4}{9}-\tfrac{4kT}{\Delta_1})\exp(-\tfrac{\Delta_1}{kT}) 
 + (\tfrac{4}{\Delta_1} + \tfrac{5}{2\Delta_2})kT - 
 \tfrac{5kT}{2\Delta_2}\exp(-\tfrac{\Delta_2}{kT})}
 {\exp(-\tfrac{\Delta_1}{kT}) + 1 + \exp(-\tfrac{\Delta_2}{kT})} \; ,
\end{equation}

\noindent while the average effective magnetic moment waves

\begin{equation} \label{mu-avr}
 \bar{\mu} = \sqrt{\frac{\mu_{\mathrm{eff},z}^2 + 2\mu_{\mathrm{eff},x}^2}{3}} .
\end{equation}

The temperature dependence of the effective moments is compared on Fig.~\ref{FigMueff} with its experimental derivation:

\begin{equation} \label{mu-avr}
\mu_\mathrm{eff}=\sqrt{\frac{3k\chi V(T-T_\mathrm{CW})}{\mu_\mathrm{0}}}
\end{equation}

\noindent where $V$ is the volume per formula unit. The Curie-Weiss temperature $T_\mathrm{CW}$ was assumed isotropic and taken at its 
$c$-axis value $T_\mathrm{CW}=+0.3\,$K. The DFT calculations and experimental data agree on a reinforcement of the anisotropy of the effective moment upon cooling. It is interesting to notice that the effective moment in the $xy$ plane  remains finite in the experimental data with $\mu_{\mathrm{eff},x}\approx 0.6\,\mu_\mathrm{B}$ whereas it fully collapses upon cooling in the DFT results predicting uniaxial magnetic moments at the $T=0$ limit.

\section{Theoretical Model Analysis and DMRG Calculation Details}

The presence of horizontal and vertical mirror planes in the crystal structure of CeMgAl$_{11}$O$_{19}$ cancels off-diagonal symmetric exchange and constrains the Dzyaloshinskii-Moriya (DM) interactions to a DM vector along the $c$ axis~\cite{Bastien2024, Gao2024}. The magnetic Hamiltonian of CeMgAl$_{11}$O$_{19}$ on the level of the nearest-neighbors interactions can be written as

\begin{equation}
H=\sum_{<i,j>}{J_\perp(S_i^xS_j^x+S_i^yS_j^y)+J_zS_i^zS_j^z+J_c(\bm{S}_i.\bm{e}_{ij})(\bm{S}_j.\bm{e}_{ij})+D_{ij}(S_i^xS_j^y-S_i^yS_j^x)} - \bm{h}\sum_i\bm{S}_i. 
\label{eq:Hsup}
\end{equation}

The first two terms of this Hamiltonian represent the bare XXZ model. The third term incorporates the effect of the bound-dependent magnetic interaction with strength $J_c$ and the fourth term the possible Dzyaloshinskii-Moriya interaction. The last term describes the influence of the external magnetic field. 

Because the typical effects expected for significant Dzyaloshinskii-Moriya interaction have not been experimentally observed ~\cite{Gao2024}, we neglect this term from further consideration and also neglect $J_c=0$ as this minimal model fits well our experimental data. An analysis of a case with finite $J_c$ can be found in Reference~\cite{Gao2024}. 

\subsection{Details of DMRG calculations}

The Density Matrix Renormalization Group (DMRG) calculations were performed using the TeNPy library (version 1.0.4)~\cite{tenpy2024}. The components of the $g-$tensor were taken at the experimental values $g_c=3.36$ and $g_{ab} \approx 0.6$. For the sake of reproducibility we state here typical parameter settings that have been sufficient for a converged result in the relevant regime for zero as well as finite magnetic field.

\begin{lstlisting}
model_params = {
        'lattice': Triangular,  # Triangular lattice
        'Lx': 12,               # Size in x-direction
        'Ly': 9,                # Size in y-direction
        'S': 0.5,               # Spin-1/2 (Heisenberg model)
        'Jx': 1.0,              # Exchange interaction Jx
        'Jy': 1.0,              # Exchange interaction Jy =Jx
        'Jz': -0.22,            # Exchange interaction Jz
        'hz': 0.5,              # Magnetic field along z-axis
        'hx': 0.0,              # Magnetic field along z-axis   
        'bc_MPS': 'finite',     # Boundary conditions for MPS
        'bc_x': 'open',         # Boundary conditions in x direction    
        'bc_y': 'cylinder',     # Boundary conditions in y direction
        'conserve':None,        # 
                }
\end{lstlisting}

\begin{lstlisting}
dmrg_params = {
        'mixer': True,
        'mixer_params': {
            'amplitude': 1.e-6,
            'decay': 1.,
            'disable_after': 18,
                        },
        'max_sweeps': 60,
        'min_sweeps': 25,
        'chi_list': {
        0: 10,
        2: 20,
        4: 40,
        8: 100,
        14: 200,
        20: 350,
                    },
            }
\end{lstlisting}
Detailed explanations of the parameters used can be found at \href{https://tenpy.readthedocs.io/en/latest/}{tenpy.readthedocs.io}. We have tested several initial states, including random states and artificially constructed antiferromagnetic configurations with spins aligned along the $z$ axis. Among these, the fastest convergence was typically achieved using the three-coloring states introduced in Ref.~\cite{Pal2021} (see Eq.~(11) in the cited work), combined with a transfer learning strategy. Specifically, the ground state at a given parameter value $x$ (e.g., $x = h_z$) obtained via direct calculation was compared to the result of a DMRG simulation initialized with a ground state computed at $x \pm \Delta x$.

\subsection{Finite size scaling}

\begin{figure}[h!] 
\begin{center} 
\includegraphics[width=0.8\linewidth]{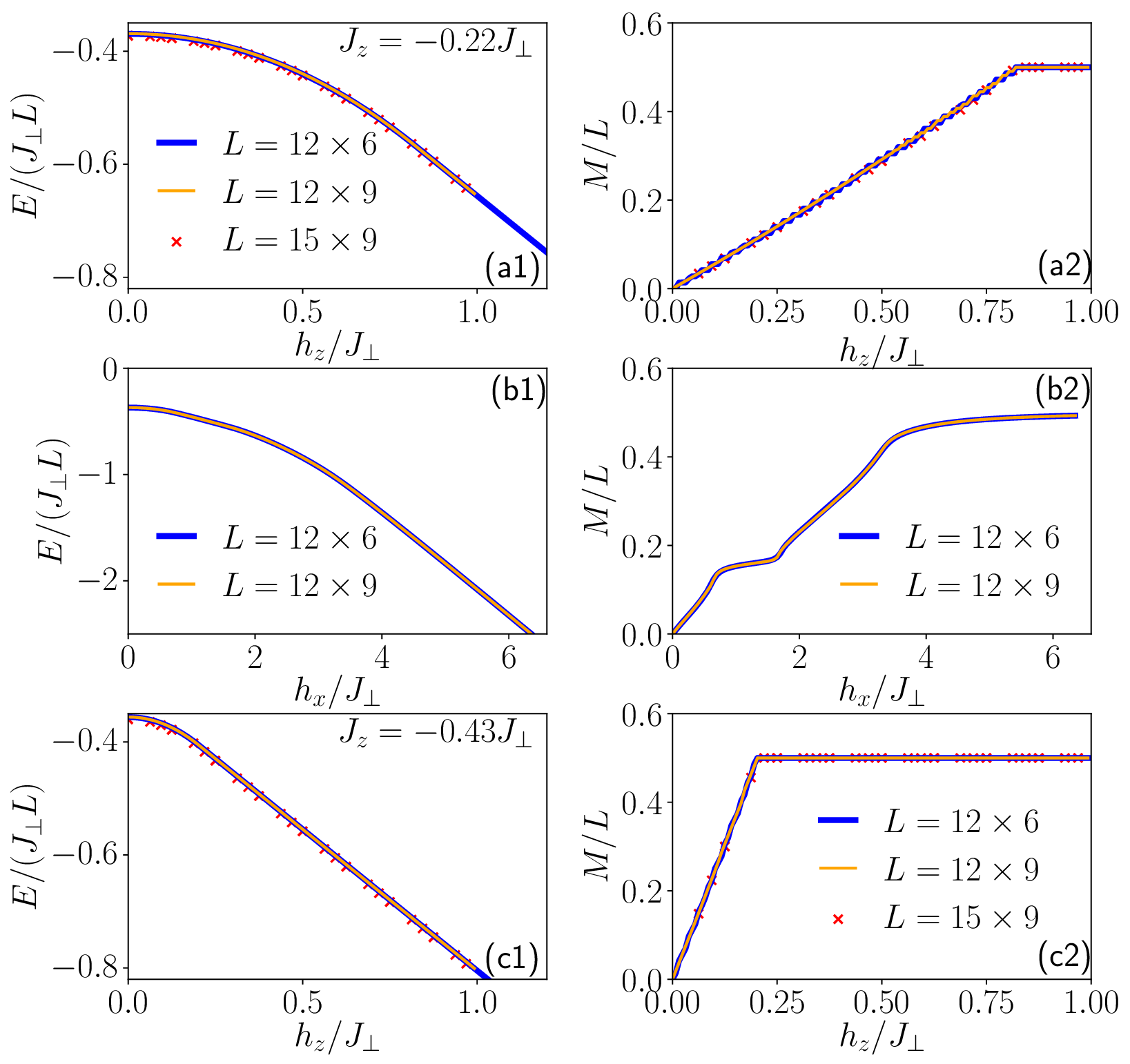} 
\caption{Comparison of DMRG results for energy (first column) and magnetization (second column) across different lattice sizes. (a) $h_z$ dependence for $J_z/J_\perp = -0.22$, (b) $h_x$ dependence for $J_z/J_\perp = -0.22$, (c) $h_z$ dependence for $J_z/J_\perp = -0.43$. Cylindrical boundary conditions were used in all cases.} 
\label{fig:DMRG_app} 
\end{center} 
\end{figure}
Since we are employing the "finite" DMRG algorithm, it is important to assess the impact of finite-size effects on our results. To this end, we have considered three lattice geometries: $L = 12 \times 6$, $12 \times 9$, and $15 \times 9$, with cylindrical boundary conditions (open in the $x$ direction and periodic in the $y$ direction). Fig.~\ref{fig:DMRG_app} displays the evolution of the energy and magnetization for $J_z = -0.22J_\perp$ and $J_z = -0.43J_\perp$ as functions of $h_z$ or $h_x$. The strong agreement across different lattice sizes suggests that finite-size effects on the magnetization are negligible.

\subsection{Structure factors}

\begin{figure}[h!]
\begin{center}
\includegraphics[width=1.00\linewidth]{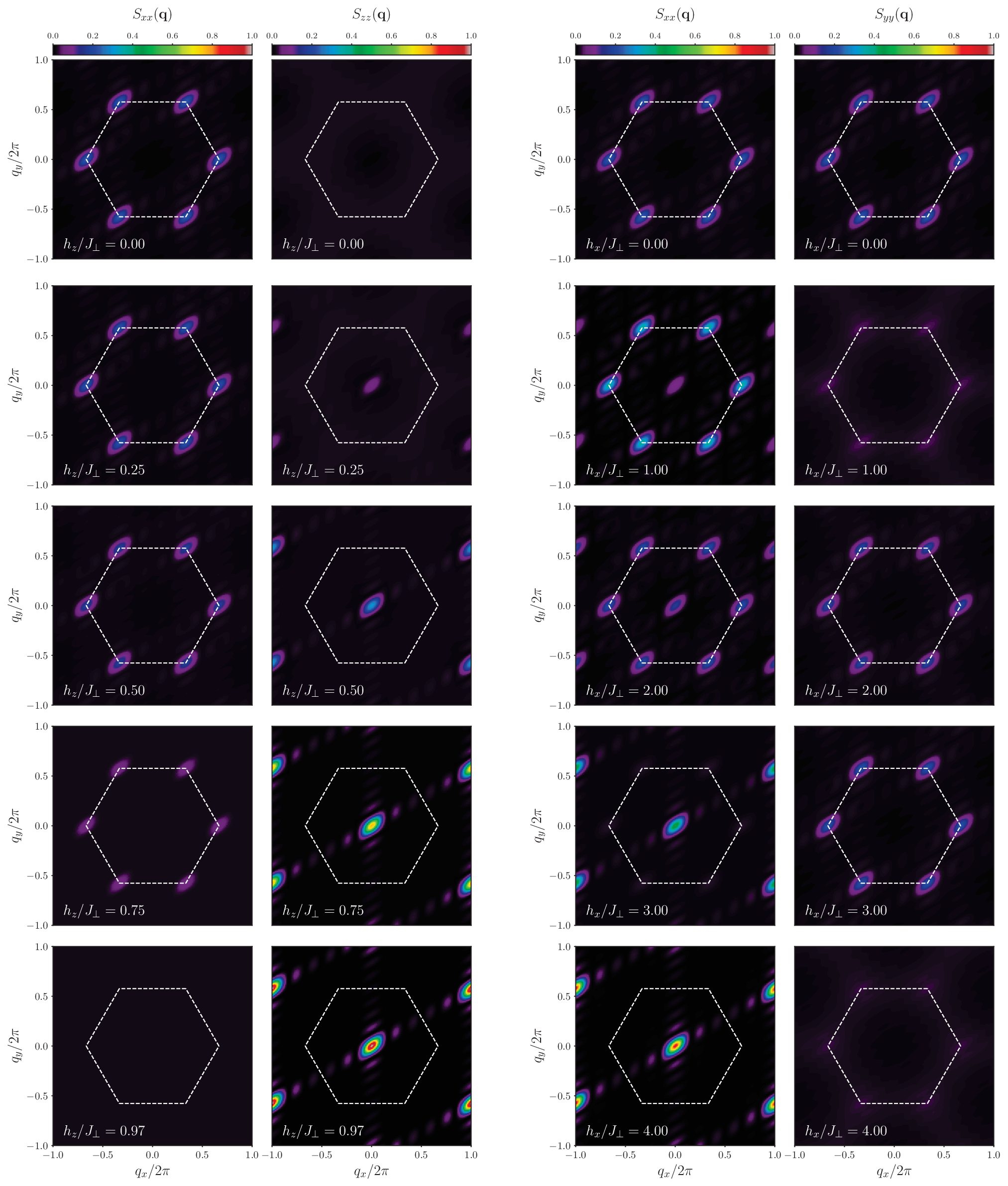}
\caption{Left columns: Spin structure factor components $S_{xx}$ and $S_{zz}$ for $J_z=-0.22J_\perp$ and increasing magnetic field (from top to bottom) $h_z$. Right columns: Spin structure factor components $S_{xx}$ and $S_{yy}$ in increasing magnetic field $h_x$.
}
    \label{fig:sf} 
\end{center}
\end{figure}
To gain deeper insight into the evolution of the magnetic ground state under increasing magnetic field in different directions, we have analyzed the components of the magnetic structure factors for :

\begin{eqnarray}
S_{xx}(\mathbf{q}) &=& \frac{4}{N^2}\sum_{m,n}e^{-i\mathbf{q} \cdot (\textbf{r}_m-\textbf{r}_n)} \langle S^x_mS^x_n\rangle,\nonumber \\
S_{yy}(\mathbf{q}) &=& \frac{4}{N^2}\sum_{m,n}e^{-i\mathbf{q} \cdot (\textbf{r}_m-\textbf{r}_n)} \langle S^y_mS^y_n\rangle,\nonumber \\
S_{zz}(\mathbf{q}) &=& \frac{4}{N^2}\sum_{m,n}e^{-i\mathbf{q} \cdot (\textbf{r}_m-\textbf{r}_n)}\langle S^z_mS^z_n\rangle,
\label{eqn:SF}
\end{eqnarray}
as shown in Fig.~\ref{fig:sf}. There the left site columns show the structure factor components $S_{xx}$ and $S_{zz}$ for an increasing magnetic field (from top to bottom) $h_z$ in the $z$ direction ($h_x=0$). The structure factors clearly show a transition from a coplanar $120^\circ$ ordering at small $h_z$ to a Ferromagnetic ordering at $h_z\approx J_\perp$. The right site then illustrates the change of ordering with increasing magnetic field $h_x$ ($h_z=0$) in the components $S_{xx}$ and $S_{yy}$. However, here we can clearly see a formation of intermediate ordering in the range of $h_{x}$ where a plateau-like structure is observed in the magnetization. The structure factor reveals that it is an up-up-down (UUD) ordering along the magnetic field direction. To make these assignments, we compared the DMRG structure-factor components with the patterns expected for the classical ground-state manifold~\cite{Starykh2015}, as sketched in Fig.~3(b) of the main text. Four field ranges emerge. From zero field up to roughly 1.1\,T, the familiar 120$^{\circ}$ coplanar order evolves into the quantum UUD state, defining the coplanar Y phase. The collinear UUD phase then remains stable up to about 2.6\,T. Beyond this field, the system enters a coplanar V phase, which persists until it converts to fully ferromagnetic order at approximately 6\,T.

\pagebreak

\bibliography{CMAO-PRL}


\end{document}